\documentclass[prl,twocolumn,aps]{revtex4}
\usepackage{graphicx}
\begin{document}

\title{Competing Orders and Spin-Density-Wave Instability in
La(O$_{1-x}$F$_x$)FeAs}

\author{J. Dong}
\author{H. J. Zhang}
\author{G. Xu}
\author{Z. Li}
\author{G. Li}
\author{W. Z. Hu}
\author{D. Wu}
\author{G. F. Chen}
\author{X. Dai}
\author{J. L. Luo}
\author{Z. Fang}
\email{zfang@aphy.iphy.ac.cn}%
\author{N. L. Wang}
\email{nlwang@aphy.iphy.ac.cn}%

\affiliation{Beijing National Laboratory for Condensed Matter
Physics, Institute of Physics, Chinese Academy of Sciences,
Beijing 100080, People's Republic of China}
%


\maketitle

\textbf{The interplay between different ordered phases, such as
superconducting, charge or spin ordered phases, is of central interest
in condensed matter physics. The very recent discovery of
superconductivity with a remarkable T$_c$= 26 K in Fe-based
oxypnictide La(O$_{1-x}$F$_x$)FeAs is a surprise to the scientific
community\cite{Kamihara08}. The pure LaOFeAs itself is not
superconducting but shows an anomaly near 150 K in both resistivity
and dc magnetic susceptibility.  Here we provide combined experimental
and theoretical evidences showing that the anomaly is caused by the
spin-density-wave (SDW) instability, and electron-doping by F
suppresses the SDW instability and recovers the superconductivity.
Therefore, the La(O$_{1-x}$F$_x$)FeAs offers an exciting new system
showing competing orders in layered compounds.}

Both charge-density wave (CDW) and spin-density-wave (SDW)
instabilities may develop in the presence of Fermi surface (FS)
nesting\cite{Gruner}, as shown in many transition metal compounds,
such as 2H-NbSe$_2$\cite{Straub} and Cr\cite{Fawcett}.  The difference
is that CDW couples to lattice and SDW couples to spin.  The CDW or
SDW instabilities may compete with other possible orderings, and
complicated phase diagrams are often drawn due to their
interplay. This is exactly what happens for LaOFeAs as shown in this
paper. LaOFeAs crystalizes in layered square lattice with Fe layers
sandwiched by two As layers (up and down), each Fe is coordinated by
As tetrahedron. Its electronic properties are dominated by the
(FeAs)-triple-layers, which contribute mostly to the electronic state
around Fermi level ($E_f$). The electronic structure of LaOFeAs is
quasi-two-dimensional and very similar to typical semi-metal, namely
there are hole-like FS cylinders around the $\Gamma$-Z line of the
Brillouin zone (BZ), and electron-like FS cylinders around the M-A
line of the BZ (except a small 3D FS around the Z point of
BZ)~\cite{Leb07,Singh,fang,Kotliar,Mazin}. We will show here that
surprisingly strong FS nesting exists by connecting the hole and
electron FS by a commensurate vector $q$=($\pi$, $\pi$, 0). This leads
to SDW instability, and it is the main cause of the anomaly observed
at 150 K experimentally. The spontaneously symmetry-broken SDW state
is characterized in terms of reduced carrier density due to (partial)
FS nesting, enhanced conductivity due to the reduction of scattering
channel, and loss of 4-fold rotational symmetry with negligible change
of lattice.

Series of La(O$_{1-x}$F$_x$)FeAs samples are fabricated, and Fig.
1 (a) shows the temperature dependence of the resistivity. The
pure LaOFeAs sample has rather high dc resistivity value and very
weak temperature dependence at high temperature, but below roughly
150 K, the resistivity drops steeply, with an upturn at lower
temperature (below 50 K). At 2$\%$ F-doping, the overall
resistivity decreases and the 150 K anomaly shifts to the lower
temperature and becomes less pronounced. At 3$\%$ F-doping, the
anomaly could not be seen, and a superconducting transition occurs
at T$_c$=17 K. With further increasing F content, and the
superconducting transition temperature increases with the highest
T$_c$=28 K seen at 6$\sim$8$\%$ F-doping. A T$_c$ vs. F-content
phase diagram is plotted in the inset of the figure 1 (a).

\begin{figure}[b]
\centerline{\includegraphics[width=3.2in]{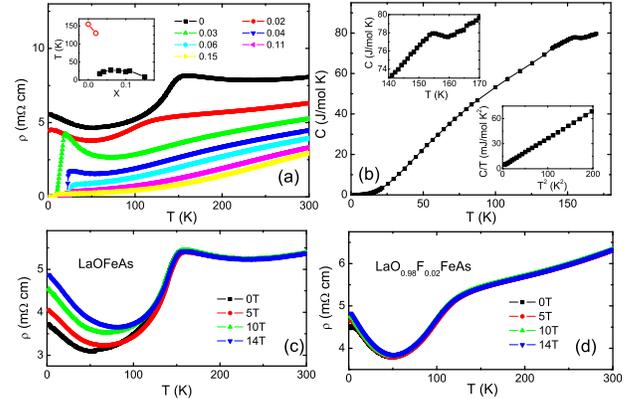}}%
\caption{(a) The electrical resistivity vs temperature for a
series of LaO$_{1-x}$F$_{x}$FeAs. Inset: The phase diagram showing
the anomaly (red cricle) and superconducting transition (black
square) temperatures as a function of F content. (b) specific heat
vs temperature curve. Upper inset: the expanded region near the
transition temperature; lower inset: a plot of C/T vs T$^2$. (c)
and (d) T-dependent resistivity under magnetic field for the pure
and 2$\%$ F-doping samples, respectively.}
\end{figure}

Apparently, superconductivity competes with the phase showing the
anomaly in this system. To identify whether the anomaly at 150 K in
LaOFeAs is due to a phase transition, we performed the specific heat
measurement. Figure 1 (b) shows the specific heat data as a function
of temperature. Very clear specific heat jump is seen at the
temperature close to 155 K (see the expanded plot in the upper
inset). A step-like, although broad, feature in the specific heat data
suggests that the anomaly is caused by a second order phase
transition. A good linear T$^2$ dependence indicates that the specific
heat C is mainly contributed by electrons and phonons.  The fit yields
the electronic coefficient $\gamma$ =3.7 mJ/mol$\cdot$K$^2$ and the
Debye temperature $\theta_D$=282 K.  Note that the electronic
coefficient $\gamma$ is significantly smaller than the values obtained
from the band structure calculations being about 5.5-6.5
mJ/mol$\cdot$K$^2$.\cite{Singh,fang} This result is unconventional,
because usually the band structure calculation gives smaller value
than the experimental data, their difference is ascribed to the
renormalization effect. We recently measured a Ni-based oxypnictide
La(O$_{0.9}$F$_{0.1}$)NiAs, and indeed found that the experimental
electronic coefficient is larger than the band structure
calculation\cite{Li}. As we shall show below that a partial energy gap
formation is revealed at the phase transition and explained as
originated from the SDW instability, the smaller experimental value
here could be naturally accounted for by the gap formation which
removes parts of the density of states below the phase transition.

\begin{figure}[t]
\centerline{\includegraphics[width=3.0in]{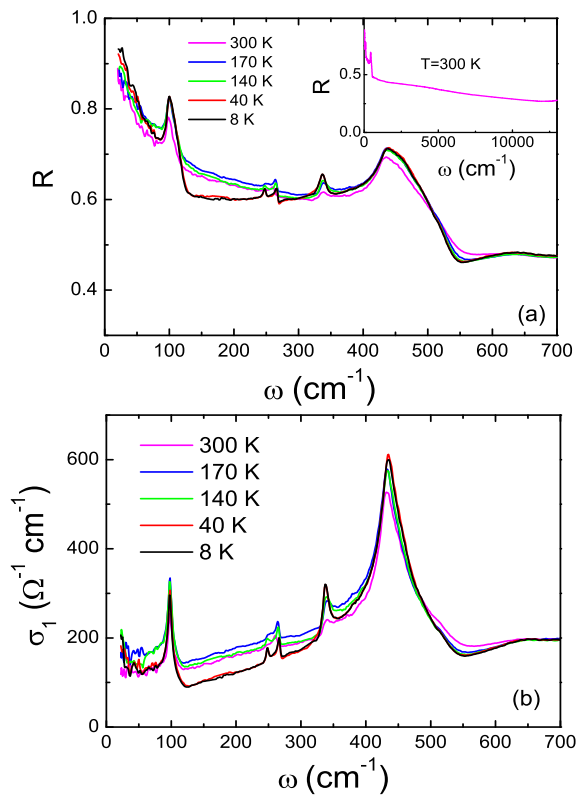}}%
\caption{(Color online) (a) The reflectance spectra in the
far-infrared region at different temperatures for the pure LaOFeAs
sample. The inset shows the reflectance over broad frequency range
at room temperature. (b) The conductivity spectra at different
temperatures.}
\end{figure}

We also measured the resistivity under magnetic field. Figure 1 (c)
and (d) shows the results for the pure and 2$\%$ F-doping samples,
respectively. The anomaly transition temperature is rather insensitive
to magnetic field up to 14 T, however sizable positive
magnetoresistance are observed at low temperature for pure LaOFeAs.
The positive magnetoresistance could be understood in terms of the
suppression of SDW order (and therefore enhanced spin scattering) by
external magnetic field.  After 2$\%$ F-doping, the low temperature
magnetoresistance becomes much weaker, which suggests the same
tendency that F-doping tries to suppress the anomaly around 150K. The
overall picture is pretty much similar to the elemental Cr, a typical
SDW system, where transition temperature is extremely insensitive to
magnetic field and sizable magnetoresistance exists after SDW
transition.\cite{Fawcett}

Important information about the nature of the phase transition could
be obtained from the optical measurement. Figure 2 (a) shows the
temperature-dependence of the reflectance in the far-infrared region
for the pure LaOFeAs sample. The reflectance at 300 K over broad
frequency range is shown in the upper inset. Even though the
measurement was performed on polycrystalline samples, two important
experimental findings regarding to the phase transition could be
unambiguously drawn from the measurement. First, below the phase
transition temperature, the reflectance is strongly suppressed below
the frequency of 400 cm$^{-1}$. This is dramatically different from
the F-doped LaO$_{0.9}$F$_{0.1-\delta}$FeAs superconducting sample
where a monotonous increase of the reflectance was
seen\cite{Chen}. The suppression is a strong indication for the
formation of an energy gap in the density of state. Note that the
reflectance at very low frequency increases fast and exceeds the
values at high temperature, e.g. above the phase transition. This
indicates that the compound is still metallic even below the phase
transition, being consistent with the dc resistivity measurement which
reveals an enhanced conductivity. The data indicate clearly that only
partial or some of the Fermi surfaces are gapped. We note that the gap
is very small, roughly in the range of 150$\sim$350 cm$^{-1}$. Below
50 cm$^{-1}$, we find a sharp upturn for conductivity spectrum at 8 K
(see fig. 2 (b)), suggesting development of very narrow Drude
compound, which should be linked to the survived FS in the SDW state.

Second, a number of pronounced phonon structures are seen in the
reflectance and conductivity spectra. Note that infrared spectroscopy
probes phonon modes only near $\Gamma$ point, the frequencies of the
observed infrared active modes are basically in agreement with the
calculations by Singh and Du.\cite{Singh} Upon cooling the sample
below the transition temperature, no new phonon mode is
observed.\cite{Remark1} Considering the polycrystalline nature of the
sample which has random orientations of the ab-plane and the c-axis,
the absence of the any new phonon mode strongly suggests that there is
no, or at least negligible, lattice distortion across the phase
transition. Therefore, the phase transition could not be of CDW
origin.

\begin{figure}
\includegraphics[clip,scale=0.2]{fig3a.eps}  \  \  \
\includegraphics[clip,scale=0.2]{fig3b.eps}
\includegraphics[clip,scale=0.25]{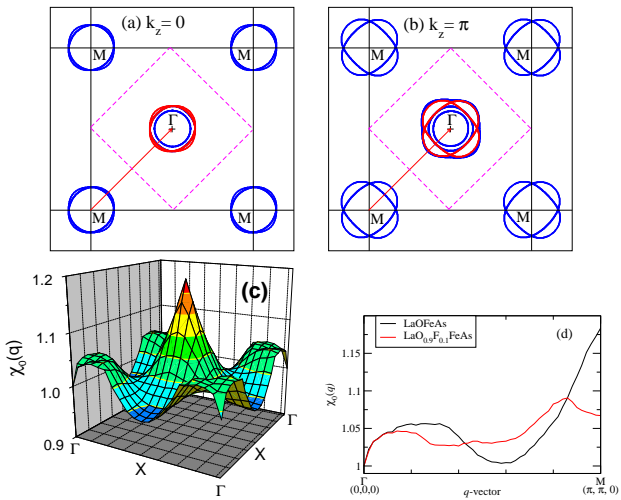} \  \
\includegraphics[clip,scale=0.25]{fig3d.eps}
\caption{(a) and (b) are the Fermi surfaces (FS) cutted at the $k_z$=0
and $\pi$ planes respectively. The blue lines are original calculated
FS, while red lines indicate the FS shifted by $q$=($\pi$, $\pi$,
0). Significant nesting is clear. (c) The calculated Lindhard response
functionn $\chi_0(q)$. It is strongly peaked at M point. (d) The
$\chi_0(q)$ along the $\Gamma$-M line. The peak at M point is much
suppressed by F-doping.}
\end{figure}

In the following we shall illustrate that the transition is caused by
the SDW transition due to the Fermi surface nesting from
first-principles calculations.  Since only the magnetic translational
symmetry has been broken, there is no modulation for the charge
density, which can explain the absence of new phonon modes appearing
in the optical data after the transition.  As we will show below,
after the SDW transtion, a stripe like spin ordering pattern appears
which breaks the 4-fold rotational symmetry and induces a nematic oder
in the charge sector.

Using the experimental high temperature structure ($P$4/nmm symmetry),
our calculated electronic structures and Fermi surfaces are identical
to those done by other people~\cite{Leb07,Singh,fang}. By cutting the
BZ into fixed-$k_z$ planes, circle like FS are resolved (as shown in
Fig.3(a) and (b) for $k_z$=0 and $k_z$=$\pi$ planes
respectively). Surprisingly, by shifting the circles around the M
points to the $\Gamma$ point, i.e. by a vector $q$=($\pi$, $\pi$, 0),
the electron-like FS will largely overlap with the hole-like FS,
suggesting significant nesting effect. For example, for the $k_z$=0
plane, the small electron circle almost exactly overlaps with the
large hole circle after shifting by $q$. Similar thing happens for the
$k_z$=$\pi$ plane. The nesting effect can be quantitatively estimated
by calculating the Lindhard response function as shown in Fig 3(c) and
(d). The calculated $\chi_0(q)$ is strongly peaked at M point for
undoped compound, and it is much suppress by electron-doping, because
the up-shift of Fermi level tends to reduce the size of hole-like FS
and enlarge the electron FS. The existence of strong nesting effect
would suggest that certain kinds of ordering, either CDW or SDW, may
develop at low temperature. As mentioned above, the absence of any new
phonon mode after the transition suggests that CDW modulation is
unlikely, otherwise the structural distortion would lead to much more
new phonon modes. Nevertheless, here we provide further theoretical
evidences.

\begin{figure}
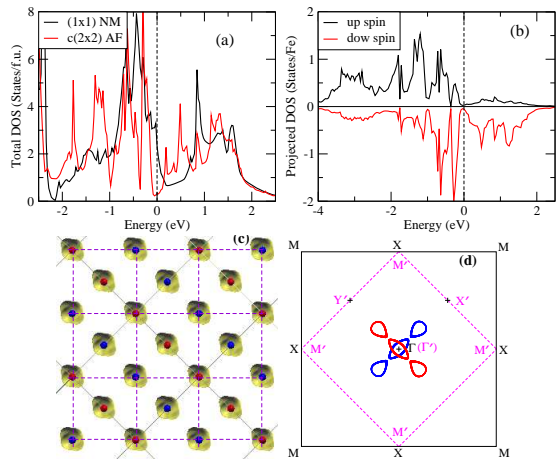

\includegraphics[clip,scale=0.3]{fig4a.eps}
\includegraphics[clip,scale=0.3]{fig4b.eps}
\includegraphics[clip,scale=0.3]{fig4c.eps}
\includegraphics[clip,scale=0.16]{fig4d.eps}
\caption{(a) The total DOS of (1$\times$1) NM and
($\sqrt{2}\times\sqrt{2}$) AF solutions. The reduction of
N($E_F$) is about 7. (b) The projected DOS of Fe site for the AF
solution. (c) The stripe-like ordering pattern of Fe plane and the
charge distribution (occupied). The Fe atoms are indicated by red
and blue spheres with up and down spin respectively. The original
(1$\times$1) unit cell is indicated by the dashed lines.  (d) The
calculated FS of AF state. The BZ special points of
($\sqrt{2}\times\sqrt{2}$) cell are indicated by symbols with
prime. The red lines are rotated from blue lines, corresponding to
the 90 degrees rotation of stripe orientation.}
\end{figure}

The commensurate nesting vector suggests the doubling of the unit cell
to be ($\sqrt{2}\times\sqrt{2}$) (i.e. the folding of the BZ along the
dashed lines shown in Fig. 3 (a) and (b)), which is not treated by
previous calculations~\cite{Singh,fang}.  We have tried to optimized
the structure of non-magnetic (NM) state without using any symmetry,
the solution always recover to the original (1$\times$1)
structure. However, among several possible magnetic solutions, one
antiferrromagnetic (AF) ordered state (see the ordering pattern shown
in Fig.4) can be stabilized with about 40 meV/Fe lowering in energy
and about 1.5 $\mu_B$/Fe for the spin moment compared to (1$\times$1)
NM solution. Clearly the nesting instability likes to couple with spin
rather charge. It is important to note that only part of the FS are
nested, as the results, the system remains to be metallic and only
partial gap opens at $E_F$. The density of state at $E_F$ is reduced
significantly compared with original (1$\times$1) NM solution as shown
in Fig. 4. All those factors are exactly what were observed in our
experiments. Treating the F-doping by virtue crystal approximation, we
found that both nesting effect and the stabilization energy of AF
solution are suppressed, again explain the experimental observation.

The obtained AF ordered state is different with the (1$\times$1) NM
state in the sense that 4-fold rotational symmetry is broken due to
the presence of magnetic ordering, and the stripe-like ordering
pattern is revealed for the Fe plane (as shown in Fig. 4
(c)). However, if the stripe orientation is rotated by 90 degrees for
the neighboring Fe plane stacking along $z$, the lattice parameters
may remain to be 4-fold invariant. Furthermore, if we concentrate on
the charge (rather than magnetic density) distribution, the
translational symmetry of total charge remains to be the same as
original (1$\times$1) structure (as shown in Fig. 4(c). As we
mentioned, this is the reason why no new phonon modes are observed. It
deserves to remark that, in the ordered state, the charge at Fe site
has a preferentially aligned distribution along stripe direction (see
Fig. 4 (c)). The fact that charge density picks up of preferred
direction after a spontaneous symmetry-breaking in spin sector could
be considered as a formation of nematic order.\cite{Fradkin} Here, the
order is characterized by a 4-fold rotational symmetry broken.

\textbf{Experiment:}

The samples were prepared by the solid state reaction using
Fe$_2$O$_3$, Fe, La, As, and LaF$_{3}$ as starting materials. LaAs was
obtained by reacting La chips and As pieces at 500 $^{\circ}C$ for 15
hours and then 850 $^{\circ}C$ for 2 hours. The raw materials were
thoroughly mixed and pressed into pellets. The pellets were wrapped
with Ta foil and sealed in an evacuate quartz tube under argon
atmosphere. They were then annealed at 1150 $^{\circ}C$ for 50
hours. The phase purity was checked by a powder X-ray diffraction
method using Cu K$\alpha$ radiation at room temperature. The XRD
patterns are well indexed on the basis of tetragonal ZrCuSiAs-type
structure with the space grounp P4/nmm.  The electrical resistivity
was measured by a standard 4-probe method. The specific heat
measurement was carried out using a thermal relaxation
calorimeter. The heat capacity of addenda were carefully calibrated
before measurement. All these measurements were preformed down to 1.8K
in a Physical Property Measurement System(PPMS) of Quantum Design
company. Optical reflectance measurement were performed on Bruker 113v
and 66v/s spectrometers in the frequency range from 20 $cm^{-1}$ to
15,000 $cm^{-1}$ at different temperatures. The samples were polished
and shinny and metallic bright surface was obtained. An
\textit{in-situ} gold overcoating technique was used for the
experiment. Optical conductivity was derived from Kramers-Kronig
transformation of reflectance. The first-principles calculations were
done using plane-wave pesudopotential method and generalized gradient
approximation (GGA) for the exchange-correlation potential. For the
1$\times$1 unit cell, the calculated results are identical to those
performed by other people~\cite{Singh,fang}.

We acknowledge the support from Y. P. Wang and valuable
discussions with Y. G. Yao and T. Xiang. This work is supported by
the National Science Foundation of China, the Knowledge Innovation
Project of the Chinese Academy of Sciences, and the 973 project of
the Ministry of Science and Technology of China.


\begin{references}
\bibitem{Remark1} A phonon mode at 247 cm$^{-1}$ is very week at
high temperature but eminently enhanced at low temperature. We
have carefully checked two different samples, and concluded that
the mode exists both below and above the phase transition.
\end{references}
\end{document}